\documentclass[twocolumn,final,prx,amsmath,amssymb,groupedaddress]{revtex4-1}

\usepackage{graphicx}
\usepackage{graphics}
\usepackage{dcolumn}
\usepackage{bm}
\usepackage{color}
\usepackage{soul}
\usepackage{textcomp}
\usepackage{rotating}

\usepackage{microtype}
\usepackage[hyperindex=true,pdftitle={Dynamical quantum non-locality},
colorlinks=true,pagebackref=false,citecolor=blue,plainpages=false,pdfpagelabels,breaklinks=true,
linkcolor=blue,urlcolor=blue]{hyperref}

\begin{document}

\title{Variability of qualitative variables: A Hilbert space formulation}

\author{Juan D. Botero}
\affiliation{Grupo de Investigaci{\'o}n en F{\'i}sica Te{\'o}rica y Matem{\'a}tica Aplicada, Instituto de F{\'i}sica, 
Facultad de Ciencias Exactas y Naturales, 
Universidad de Antioquia UdeA; Calle 70 No. 52-21, Medell\'in, Colombia.}

\author{Leonardo A. Pach\'on$^*$}
\affiliation{Grupo de Investigaci{\'o}n en F{\'i}sica Te{\'o}rica y Matem{\'a}tica Aplicada, Instituto de F{\'i}sica, 
Facultad de Ciencias Exactas y Naturales, 
Universidad de Antioquia UdeA; Calle 70 No. 52-21, Medell\'in, Colombia.}

\begin{abstract}
A new formalism to express and operate on diversity measures of qualitative variables, 
built in a Hilbert space, is presented. 
The abstract character of the Hilbert space naturally incorporates the equivalence 
between qualitative variables and is utilized here to (i) represent the binary character of 
answers to categories and (ii) introduce a new criterium for choosing between different 
measures of diversity, namely, robustness against uncertainty.
The full potential of the formulation on a Hilbert space comes to play when addressing the 
reduction of categories problem, a common problem in data analysis.
The present formalism solves the problem by incorporating strategies inspired by mathematical 
and physical statistics, specifically, it makes use of the concept of partial trace.
In solving this problem, it is shown that properly normalizing the diversity measures is instrumental 
to provide a sensible interpretation of the results when the reduction 
of categories is performed. 
Finally, the approach presented here also allows for straightforwardly measuring diversity 
and performing category reduction in situations when simultaneous categories could be chosen.
\end{abstract}

\maketitle

\date{\today}

\section{Introduction}
In a wide range of research fields, such as biology \cite{Sim49,Kva91}, economics 
\cite{FOR:FOR2340}, political science \cite{Wil73}, marketing, communications 
\cite{LM98,MD03} and other social science, the statistical analysis of qualitative variables 
is fundamental to construct models that predict the behaviour of systems of interest [see, 
e.g., \cite{FOR:FOR2340,MD03,BM00,BK10}.
There are two kinds of qualitative variables, nominal and ordinal and from the statistical point 
of view, they should be manipulated differently \cite{KP07}. 
Particularly, measuring variability among a set of qualitative variables is 
different when the set is composed of ordinal or nominal variables \cite{Rey84}. 

Computing the variability of a set composed of quantitative variables is straightforward: 
compute, e.g., the standard deviation of the data to know how close or spread the data is 
distributed \cite{Fre07}. 
Conversely, for a set of qualitative variables, this is not a trivial matter and it is not clear 
how the spreading of a given dataset should be formally calculated. 
In the literature, there can be found a copious number of approaches aimed to calculate the 
variability of qualitative data \cite{Sim49,Wil73,Wal77,Blau77,AA78,Tea80,MA95,Lau98,
LM98, WO98} proposing a different kind of indices to do so.
In the last years, some efforts have also been devoted to that direction, 
e.g., based on the underlying concept of diversity, some of the mathematical 
expressions used to measure variability have been discussed and justified (cf. e.g., 
\cite{MD03,BM00,HK07}).
Some of the efforts also aimed at classifying variability indices based 
on the structure of the expressions \cite{MD03,BM00} or their type of measurement \cite{HK07}, 
such as separation variety and disparity. 
But none of them has been formulated in a basis independent manner.

Two problems related to the sensibility of the measures are addressed here.
The first is the robustness of the measures under the presence of noise or uncertainty that may 
arise, e.g., when analyzing data from a selected fraction of a given community (sample).
Hence, having a classification of the variability indices based on their 
robustness is pertinent when information is incomplete.

The second problem is the influence of the reduction of categories on the variability measurements. 
Some work in this direction has been done in the past \cite{Wil73,MD03}. 
Specifically, the proposal made in Ref.~\cite{Wil73} consists in replacing by zero the proportion of 
the category that will be reduced and then redistribute the proportions equally among the other 
categories. 
From an information science perspective, this strategy violates Landauer's
principle or equivalently, it violates the second law of thermodynamics (see below).
Alternatively, in Ref.~\cite{MD03} an approach based on performing a linear regression of 
the data was suggested, it allows for analyzing how sensible the measures are to the number 
of categories $k$ and the maximum proportion among categories $P_\mathrm{max}$. 
The Hilbert space formalism presented here leads to a straightforward solution to this problem
in terms of the partial trace method \cite{breuer2007theory}, that contrary to the previous
ones \cite{Wil73,MD03}, it does not disregard information because when the partial 
trace is applied the dimension of the resulting state is smaller than the original but keeps all
the information provided by the tracedout states.

The dimension of the proportion matrix is increased when mapped onto the Hilbert space, 
(see below), so that this representation 
enables the analysis of the variability when several categories may be selected 
simultaneously. e.g., in a poll when several answers can be chosen at the same time. 

\section{Variability measures and the density operator}
\label{Sec:vrbltymsrmnts}
Variability is a measure of how spread or localized a set of variables is, i.e., it distinguishes 
if the frequency of the variables are mainly localized in one variable or if is distributed among 
them.
The dataset can be composed of qualitative or quantitative variables, here, interest is on the 
former kind.
Four different expressions, ($\sigma_\mathrm{L}$, $\sigma_\mathrm{E}$, $\sigma_\mathrm{S}$, 
$\sigma_\mathrm{P}$), frequently found in the literature \cite{FOR:FOR2340,Wil73,AA78} are 
reformulated here to calculate the variability of qualitative data sets and to allow comparisons 
among them. 
\begin{figure}
\centering
  \includegraphics[width=0.225\textwidth]{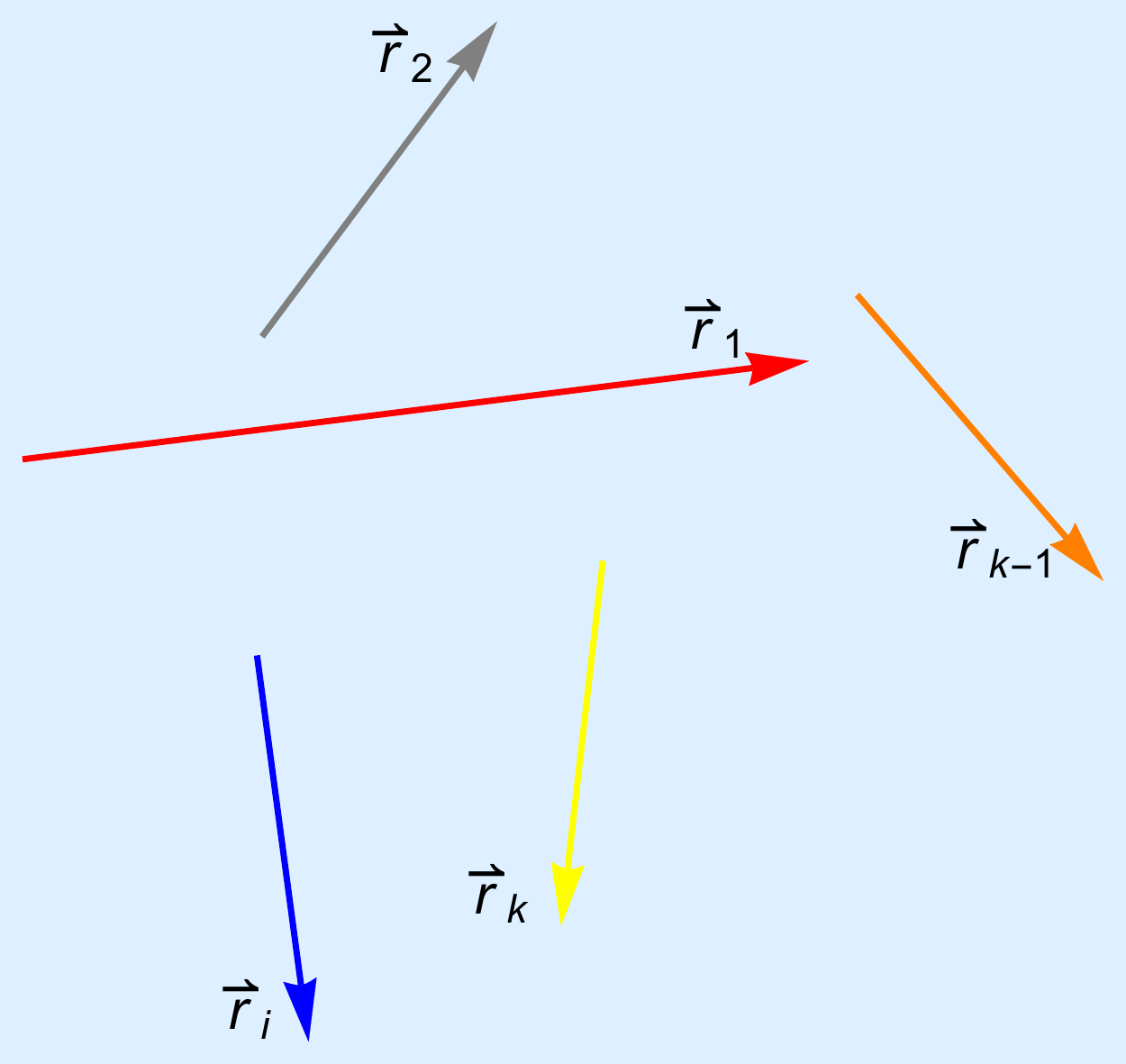} 
  \includegraphics[width=0.25\textwidth]{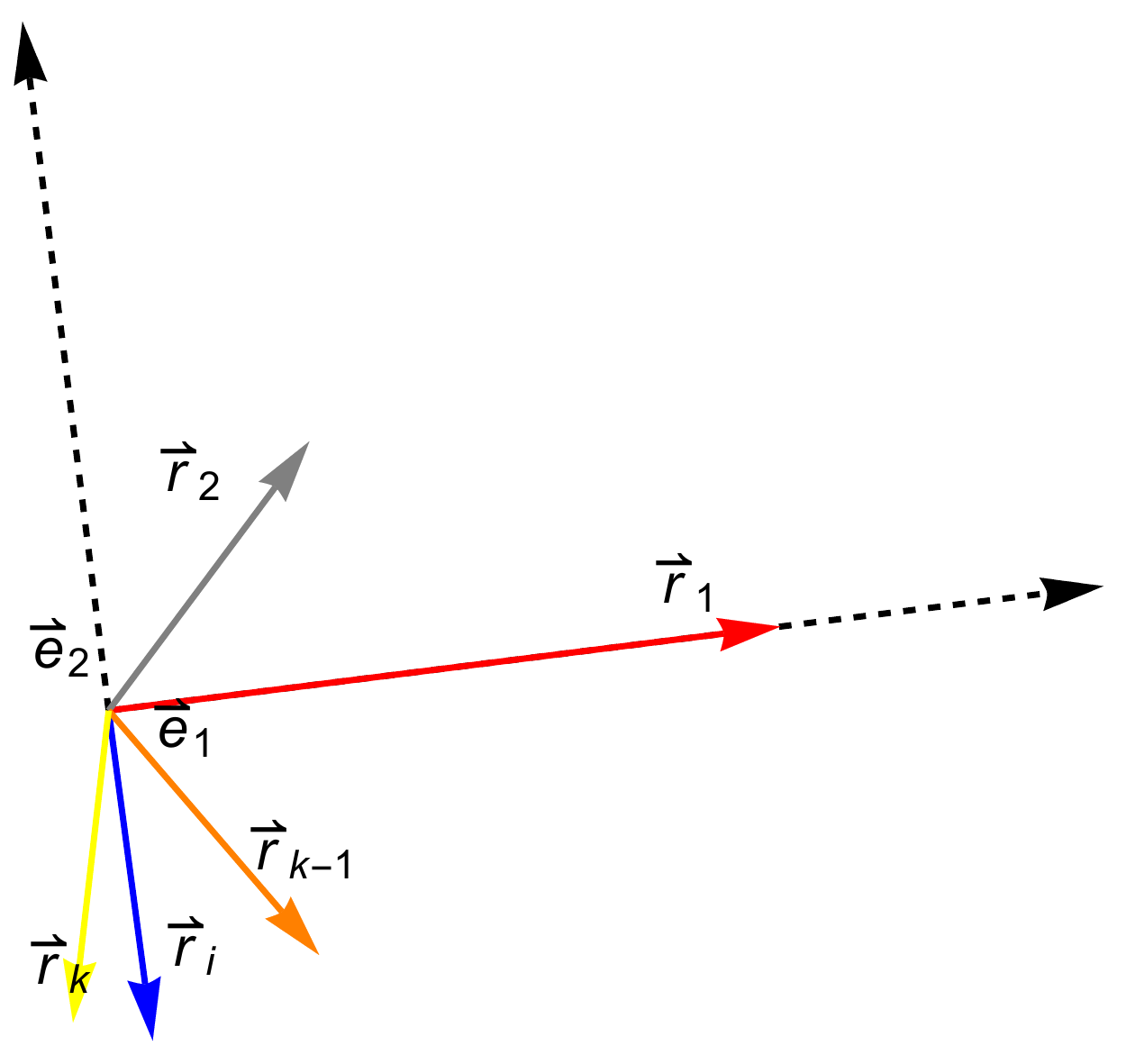} 
\caption{Left panel: $\{\vec{r}_1, \vec{r}_2, \ldots, \vec{r}_i, \ldots  
\vec{r}_{k-1}, \vec{r}_k\}$ denotes  an abstract set of elements in a linear two-dimensional vector space. 
Right panel: A particular Cartesian orthonormal basis $\{\vec{e}_1,\vec{e}_2\}$, that defines the identity 
``operator'' (dyadic) $\hat{\mathsf{1}} = \vec{e}_1\vec{e}^{\,1}+\vec{e}_2\vec{e}^{\,2}$ with 
$\vec{e}_i\cdot\vec{e}_j=\delta_{ij}$, is selected to represent the abstract set.
Thus, every element can be expressed as $\vec{r}_l = \vec{r}_l  \cdot \hat{\mathsf{1}} = 
(\vec{r}_l\cdot \vec{e}_1) \vec{e}^{\,1} + 
(\vec{r}_l \cdot \vec{e}_2) \vec{e}^{\,2}$. 
Note that any complete basis $\{\vec{\epsilon}_1,\vec{\epsilon}_2\}$ can be selected and that operations 
over the elements $\vec{r}_l$ should be independent of 
the particular selection.}
 \label{Fig:CncptlBss}
\end{figure}

A fundamental aspect towards the formulation of variability measures is the concept
of a complete basis set \cite{You88,Ber99}.
To be concrete, consider a two-dimensional linear vector space as in Fig.~\ref{Fig:CncptlBss}.
Being elements of a linear vector space, vectors $\{\vec{r}_1, \vec{r}_2, \ldots, \vec{r}_i, \ldots  
\vec{r}_{k-1}, \vec{r}_k\}$ are all equivalent and there is not preferable sorting until a particular
complete set of basis vectors $\{\vec{e}_1,\vec{e}_2\}$, with unit dyadic $\hat{\mathsf{1}} = 
\vec{e}_1\vec{e}^{\,1}+\vec{e}_2\vec{e}^{\,2}$, is selected to represent the initial 
vector set. 
The unit vectors $\vec{e}^{\,i}$ is the basis of the dual space \cite{You88,Ber99}.
This situation is certainly analog to the situation of qualitative variables.
Once the basis is selected, vectors can be sorted, e.g., depending on their projection to any of 
the basis elements, $\vec{r}_l \cdot \vec{e}_i$.
As any basis set can be utilized, operations over the elements of the space should be independent
of the basis.
A metric formulation in terms of a basis set with operations that are independent of the basis is the 
route followed here for qualitative variables.

In doing so, define $k$ as the number of categories, $n_i$ the number of answers associated to 
the $i^{\mathrm{th}}-$category, $N$ the total number of answers, i.e., $N=\sum n_i$ and 
$p_i=n_i/N$ the proportion of answers. 
The central objects towards a Hilbert space formulation are the ``density operator" $\hat{\rho}$
and the complete basis $\{| 0_1 \ldots 1_i \ldots 0_j \ldots 0_k\rangle\}$ in the Hilbert space 
$\mathcal{H}$.
Specifically, for $k$ categories, the basis element $| 0_1 \ldots 1_i \ldots 0_j \ldots 0_k\rangle$ 
represents the situation when the $i^{\mathrm{th}}$-category was selected with $p_i$=1.
Similarly, $| 0_1 \ldots 0_i \ldots 1_j \ldots 0_k\rangle$ represents the situation when the 
$j^{\mathrm{th}}$-category was selected with $p_j$=1.
Note that any other complete basis set could be selected; however, the basis set 
$\{| 0_1 \ldots 1_i \ldots 0_j \ldots 0_k\rangle\}$ intrinsically accommodates the proportion 
of answers to each category.
The identity operator reads $\hat{\mathsf{1}} = \sum_i | 0_1 \ldots 1_i \ldots 0_j \ldots 0_k\rangle
\langle 0_1 \ldots 1_i \ldots 0_j \ldots 0_k| $, where the elements $\langle 0_1 \ldots 1_i \ldots 0_j \ldots 0_k|$
are elements of the dual of the Hilbert space \cite{You88,Ber99}.

The matrix representation of the density operator $\rho=\hat{\mathsf{1}} \cdot \hat{\rho} \cdot 
\hat{\mathsf{1}}$, 
in the present formulation, corresponds to a diagonal matrix with entries $p_i$ and 
with $\sum_i^N p_i=1$ (see below).
%
%
%
Projections of the density operator onto a particular basis allows for manipulating 
qualitative and quantitative on the same ground; thus unifying operational tools.
The present formulation may be utilized to define probability distributions for qualitative variable
in the same form as probability distributions are defined for finite dimensional Hilbert spaces 
\cite{Woo87,KR&09}.
This may be of great relevance in a wide range of scientific areas such as, political science, 
marketing, sociology, biology and  economics, where the presence of qualitative variables is of great importance.
For the later, it is also important a simultaneous manipulation of both, qualitative and quantitative
variables to enhance predictability models \cite{MSY15}

In general, the matrix representation of $\hat{\rho}$ onto the states defined above considers elements 
of the type $| 0_1 \ldots 1_i \ldots 1_j \ldots 0_k\rangle$ or $| 1_1 \ldots 1_i \ldots 1_j \ldots 1_k\rangle$ 
that account for situations when several categories can be selected simultaneously, i.e., situations when 
categories $i^{\mathrm{th}}$ and $j^{\mathrm{th}}$ can simultaneously be selected (i.e., element 
$| 0_1 \ldots 1_i \ldots 1_j \ldots 0_k\rangle$) or when all the possible categories may be selected 
(i.e., element $| 1_1 \ldots 1_i \ldots 1_j \ldots 1_k\rangle$).
For simplicity and concreteness, those situations are disregard here by assuming that their corresponding
proportion is zero.
The resulting elements form a complete set for a subspace that by analogy to quantum mechanics, 
it will be referred to as the \textit{singly-excited manifold} and its basis elements as 
\textit{singly-excited states}.
After projecting the density operator $\hat{\rho}$, it takes the following form
\begin{widetext}
\begin{align}
\label{equ:densitymatrix}
\rho=
 \bordermatrix{         & | 0 ... 0 ...0\rangle & | 1 0 ...0\rangle & | 0 ... 1_i ...0\rangle & | 0 ... 0 ...1_k\rangle & \cdots & | 1 ... 1_i ...1_k\rangle \cr 
\langle 0 ... 0 ...0|   & 0                     & 0                 & 0                       & 0                       & \cdots & 0                       \cr
\:\:\langle 1 0 ...0|   & 0                     & p_1               & 0                       & 0                       & \cdots & 0                       \cr
\langle 0 ... 1_i ...0| & 0                     & 0                 & p_i                     & 0                       & \cdots & 0                       \cr
\langle 0 ... 0 ...1_k| & 0                     & 0                 & 0                       & p_k                     & \cdots & 0                       \cr
\:\:\:\:\:\:\:\vdots & \vdots                & \vdots            & \vdots                  & \vdots                  & \ddots & \vdots                  \cr
\langle 1 ... 1_i ...1_k| & 0                   & 0                 & 0                       & 0                       & 0      & 0                       \cr
}.
\end{align}
\end{widetext}
Under this consideration, $\rho$ is a diagonal matrix and only one sector 
is non-zero, namely, the one for which only one category may be selected.

Note that the present formulation resembles the One Hot Encoding Algorithm 
\cite{MG16}, widely used in Artificial Intelligence literature.
However, as it is shown below, the construction based on Hilbert spaces allows 
for making use, in social sciences, of all analysis tools form, e.g., quantum mechanics.
This fact opens the door for a more formal and precise analysis of qualitative variables.  


\subsection{Variability measures in terms of the density operator}
Once the density operator is projected onto a particular basis, the next step is to calculate
its variability in a way that is independent of the representation.
This is achieved below by introducing the trace operation, a basis invariant operation, over
the density operator $\hat{\rho}$.
In the literature, it can be found several types of variability measures and attempts to classify them 
\cite{BM00, MD03, HK07}.
In this work we are going to focus in the following four: 

\textit{Measure Type I}-- The first measure $\sigma_\mathrm{L}$ is associated with the 
linear difference between all the answers and can be written as
\begin{equation}
 \label{eq:dispLop}
 \sigma_\mathrm{L}=1-\frac{1}{2(k-1)}\sum_{\alpha=1}^k\sum_{\beta>\alpha}^k 
\mathrm{tr}\left|\hat{\rho}-\Pi_{\alpha\beta}\hat{\rho}\Pi_{\alpha\beta}^\dagger \right |,
\end{equation} 
where the matrix $\Pi_{\alpha\beta}$ represents the permutation matrix associated 
with the rows $\alpha$ and $\beta$, i.e., the matrix obtained by permuting the rows $\alpha$ 
and $\beta$ from the identity matrix $I$ of order $2^k$ \cite{Bru06}. 
The trace operation $\mathrm{tr(\cdot)}$ represents the sum of the eigenvalues of the 
matrix and $\dagger$ refers to the complex transposition.

With a different normalization and without the Hilbert space formulation, an equivalent expression 
of the Eq.~(\ref{eq:dispLop}) where mentioned before by Wilcox \cite{Wil73} as MDA, 
emphasizing that the main characteristic is that it is ``dependent on the spread of the variate-values 
among themselves and not on the deviations from some central value''.

\textit{Measure Type II}-- The second measurement is similar to the later, but the 
distance is calculated as an Euclidean distance instead and has the form
\begin{equation}
 \label{eq:dispEop}
 \sigma_\mathrm{E}=1- \frac{1}{2(k-1)}\left[\sum_{\alpha=1}^k\sum_{\beta>\alpha}^k 
 \mathrm{tr} (\hat{\rho}-\Pi_{\alpha\beta}\hat{\rho}\Pi_{\alpha\beta}^\dagger)^2\right]^{\frac{1}{2}}.
\end{equation}
An equivalent measure was proposed by Tsui \emph{et al} \cite{TEO92} in 1992 and named later in 2007 as MED
(Mean Euclidean Distance) by Harrison and Klein \cite{HK07}. This measure also has the same property of the 
Mean linear distance mentioned by Wilcox.

The following two variability expressions are frequently found and used in the qualitative statistics literature,
even though it has their ground in the physics literature.

\textit{Measure Type III}--The Shannon or von Newman entropy \cite{Sha01} 
can be directly extended to the present formalism after introducing a proper normalization 
factor. 
Traditionally, the normalization factor is taken as $1/\log_2 k$ that corresponds to the maximum 
entropy encoded in a density operator of dimension $k \times k$.
Because $\hat{\rho}$ may include multi-selection of categories, the maximum entropy measurement 
corresponds to replacing $k$ in favor of $2^k$. 
Thus, 
\begin{equation}
 \label{eq:dispSHAop}
 \sigma_\mathrm{S}=
 \frac{\mathrm{tr}(\hat{\rho} \log_2 \hat{\rho})}{\mathrm{tr}(\hat{\rho}_\mathrm{max} \log_2 \hat{\rho}_\mathrm{max})},
\end{equation}
where $\hat{\rho}_\mathrm{max}$ is the density matrix associated to the configuration of 
maximum entropy of the \textit{singly-excited manifold}, i.e., a density matrix with the entries 
on the single excited states equal to $1/k$ and the rest of them equal to zero. 

In social contexts, the above index was first introduced in the context of behavioral science by Senders 
\cite{Sen58} as a measure of uncertainty, ``which will be high when the number of alternative possibilities 
is high, and low when some of the possibilities are much more likely than others''.

\textit{Measure Type VI}-- The last measure type, also known as the Index of Qualitative 
Variation (IQV), is the most commonly used in literature ranging from psychology, politics, economy and 
others. 
Its equivalent in physics, particularly in quantum mechanics, is known as linear entropy
and corresponds to one minus the purity of the density operator $\hat{\rho}$, namely, $1-\mathrm{tr}\hat{\rho}^2$.
Specifically, $\sigma_\mathrm{P}$ takes the following form
\begin{equation}
 \label{eq:dispPop}
 \sigma_\mathrm{P}=\frac{1-\mathrm{tr} \hat{\rho}^2}{1-\mathrm{tr} \hat{\rho}_\mathrm{max}^2},
\end{equation}
where $\hat{\rho}_\mathrm{max}$ is defined as in Eq.~(\ref{eq:dispSHAop}). 
Traditionally, the normalization used for the IQV is $k/(k-1)$, which also corresponds to the maximum 
linear entropy of a density matrix with dimension $k \times k$.
In practical terms, the denominators of Eq.~(\ref{eq:dispSHAop}) and (\ref{eq:dispPop}) are equal to 
the traditional normalization factors $-1/\log_2 k$ and $k/(k-1)$, respectively, but when the reduction of 
categories is applied, the selection of a proper normalization factor becomes not trivial (see below).

\begin{figure}
\centering
\includegraphics[width=0.45\textwidth]{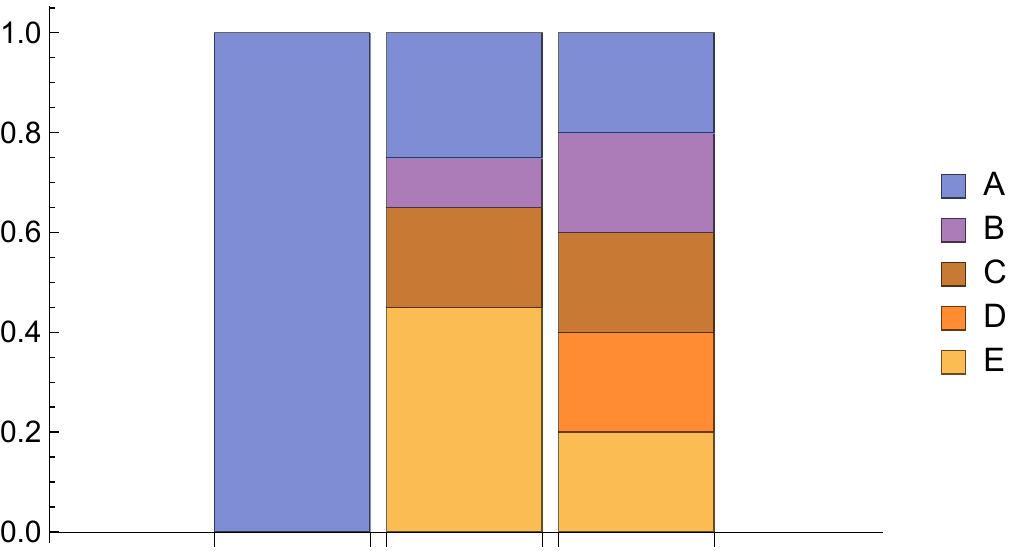}
\caption{Pictorical representation of the extreme values of the variability measurement $\sigma$ in a 
five categories escenario named $\{$A,B,C,D,E$\}$ for one hundred answers.
{\bf Left Panel} $\sigma=0$, there is no variability among 
the categories, which means all the information localized in one category. {\bf Central Panel} $0<\sigma<1$,
a situation in between is also presented, where the information is ``randomly'' spread among the categories.
{\bf Right Panel} $\sigma=1$, the variability is maximum, i.e., the information is maximally spread and equally 
distributed.}
 \label{Fig:pctrcl}
\end{figure}

All previous measures of variability are normalized to 1 so that the limits bounds as $0$ and $1$. 
The lower limit is reached when all the categories are zero except one (see Fig.~\ref{Fig:pctrcl}) 
and can be interpreted as no 
variability at all, and in terms of information, a maximum knowledge of the system. 
On the other hand, the upper limit appears when all the categories have the same value $1/k$, 
i.e., all of them are equally distributed. Having all the indices bounded by the same values, $[0,1]$, 
allows to make direct comparisons among them and also permits a better interpretation of the 
results.
All the indices presented in Eqs. (\ref{eq:dispLop}) to (\ref{eq:dispPop}), in some sense, measures 
how the bars in Fig.~\ref{Fig:pctrcl} are distributed among
the categories and also take into account the size of the bars. Interpreting the extremes values of 
$\sigma=0$ and $\sigma=1$  as completely opposite characteristics: Uncompetitive-Competitive, 
Homogeneous-Heterogeneous, Agreement-Disagreement, Segregated-Integrated and Localized-Delocalized, 
mentioning some of them suggested by Wilcox \cite{Wil73}, the variability can be understood as a 
measure of how equal or unequal is a set of categories.

\subsection{Robustness of variability measures under uncertainty}
\label{ssub:rbstnssVrblyMsrs}

All previous measure types provide an idea about how spread the categories are and despite they have 
the same bounds, their intermediate values may differ. 
Therefore, interest here is in providing a new classification criterium for selecting, when conceptually
possible \cite{MD03}, one measure over the others; specifically, the classification proposal is based 
on the robustness against uncertainty. 
Motivation for proposing this criterium is clear, when dealing with real datasets uncertainty is always 
present, e.g., (i) elasticities of commodity prices with respect to supply or demand \cite{Roe07} represents
uncertainty; (ii) the number of violent acts in a society has uncertainty due to the unregistered acts; 
(iii) when results of a poll are obtained from a sample community there is uncertainty.

In this research, uncertainty was artificially introduced by means of a stochastic variable, i.e., by adding noise to the 
proportion $p_i$.
Specifically, this is done by adding up a random number $\zeta_i$ selected from the noise domain.
That is to say, e.g., for a noise of amplitude $5\%$, a random number $\zeta_1$ is selected 
between $[-0.05,0.05]$ and added to the proportion of the first category, $p_1$; then, another 
random number $\zeta_2$ is selected from the same range and added up to the second category, 
$p_2$; the same process is repeated until the $(k-1)^{\mathrm{th}}$-category is reached. 
For the last category, $p_k$, and with aim of preserve the normalization,  $\sum_i^k p_i=1$, 
the number obtained by the summation of the all previous proportions is subtracted from 1. 
Each set of random numbers $\{\zeta_1,\zeta_2,\ldots \zeta_{k-1}\}$ form a realization 
of the random variable $\zeta$. 
Since each number is random, each realization $\zeta$ is random as well.
Therefore, to guarantee reproducibility and to simulate a more realistic situation, thousands 
of realizations are averaged until no changes above $10^{-4}$ are detected in the measures,
i.e., until convergence is reached.
The number of realization to achieve convergence varies from situation to situation and heavily
depends on the noise amplitude.
For the cases considered below, convergence was reached after averaging out over $10^4$ 
realizations.

\section{Application to Census of problem-solving courts, 2012}
\label{Sec:App}
As an application of the formalism introduced here, consider the recently released dataset
of the Problem-solving Courts, by State and Selected U.S. Territories, 2012 published by 
the Bureau of Justice Statistics \cite{SRK16}.
The problems solved by the courts in the U.S.  are classified in ten different categories, 
namely, Drug, Mental health, Family,  Youth specialty, Hybrid DWI/drug, DWI, Domestic violence, 
Veterans, Tribal wellness and Other \cite{SRK16}.
For the following analysis, each problem is considered as a category with a certain
proportion $p_i$ and variability analysis is performed by each state.

Table~\ref{Tab:problems} presents two opposite states in terms of the spreading of the 
problems-solving courts, Arkansas (Ark.) and Wisconsin (Wis.). 
It can be seen that  problems in Arkansas are mainly due to drugs and the other categories 
have not significant contribution; on the other hand, and even though the main problem 
in Wisconsin is also drugs, the other categories also have significant values. 
Hence, Arkansas is expected to have a small variability and contrary, Wisconsin should have high 
variability values. 
\begin{widetext}
\begin{center}
  \begin{table}
 \begin{tabular}{| c | c | c | c | c | c | c | c | c | c | c |}
 \hline
          &     & Mental &        & Youth     & Hybrid   &     & Domestic &          & Tribal   &      \\
          &Drug & health & Family & specialty & DWI/drug & DWI & violence & Veterans & wellness & Other \\ \hline 
 Ark. & 49 & 1 & 0 & 3 & 0 & 0 & 1 & 1 & 0 & 0\\
Wis. & 18 & 3 & 2 & 7 & 8 & 11 & 1 & 10 & 2 & 1 \\ \hline
\end{tabular}
\caption{The number of problem-solving courts for the states of Arkansas (Ark.) and Wisconsin (Wis.)}
\label{Tab:problems}
\end{table}
\end{center}
\end{widetext}
The variability values of the four measurements $\sigma_{\mathrm L}$, $\sigma_{\mathrm E}$,
$\sigma_{\mathrm S}$ and $\sigma_{\mathrm P}$ for the two states are presented in Table \ref{Tab:dispersions}. 
As anticipated, all measures have a smaller value for Arkansas than for Wisconsin, 
but the values of the measurements significantly differ among them.
Therefore, a comparison can be made among states with the same measurement, 
but not between measures within the same state.
\begin{table}
 \begin{center}
  \begin{tabular}{| c | c | c | c | c |}
\hline
          & $\sigma_\mathrm{L}$ & $\sigma_\mathrm{E}$ & $\sigma_\mathrm{S}$ & $\sigma_\mathrm{P}$\\ \hline
 Ark. & 0.0484 & 0.1195 & 0.2085 & 0.2248\\
 Wis. & 0.4973 & 0.7200 & 0.8497 & 0.9216 \\ \hline
\end{tabular}
 \end{center}
\caption{The numerical values of the four variability measurements $\sigma_\mathrm{L}$, $\sigma_\mathrm{E}$, 
$\sigma_\mathrm{S}$ and $\sigma_\mathrm{P}$ for the states of Arkansas (Ark.) and Wisconsin (Wis.)}
\label{Tab:dispersions}
\end{table}

%
%
%

\begin{figure}
\centering
  \includegraphics[width=0.45\textwidth]{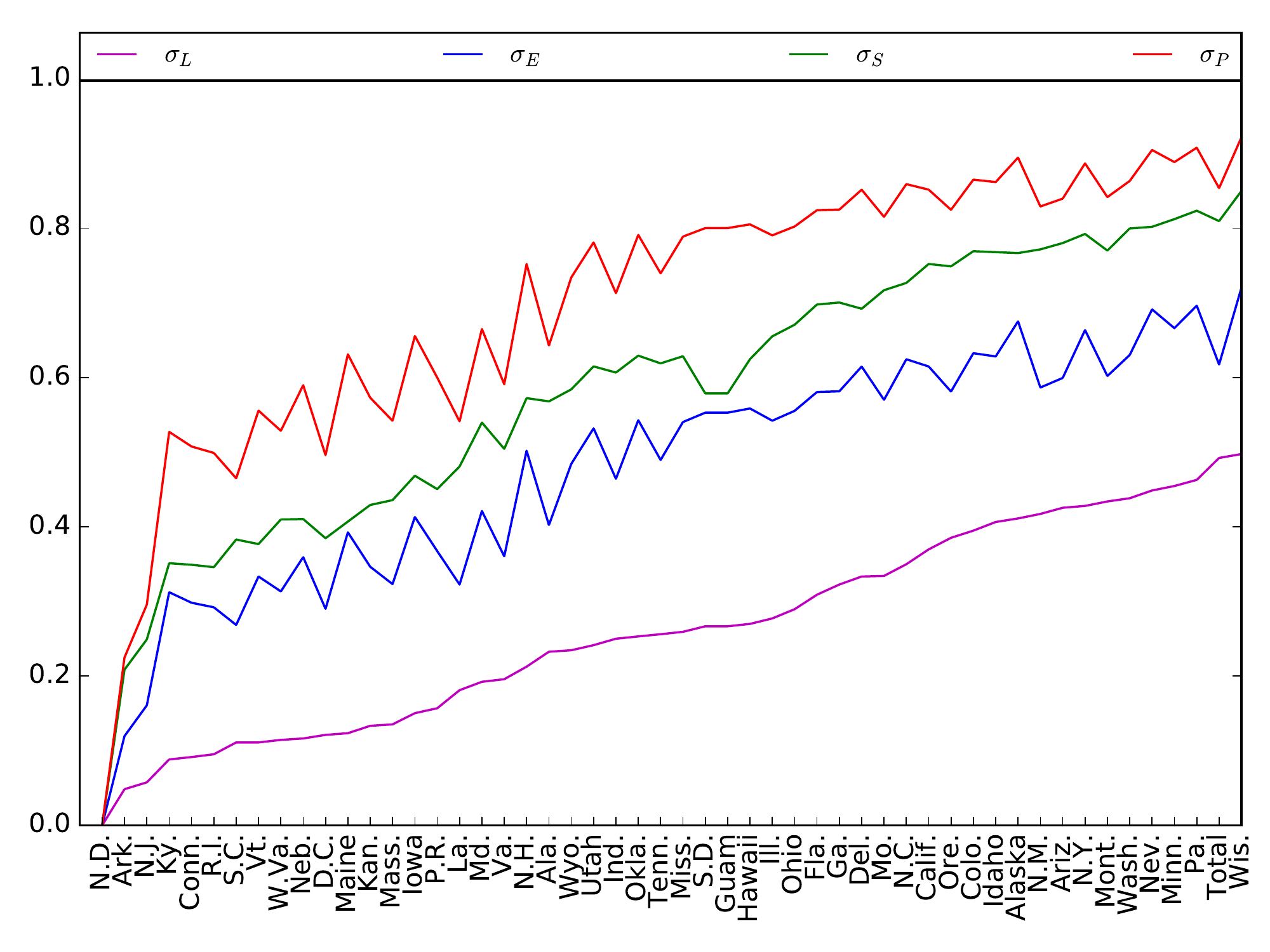} 
\caption{$\sigma_{\mathrm{L}}$, $\sigma_{\mathrm{E}}$, $\sigma_{\mathrm{S}}$ and 
$\sigma_{\mathrm{P}}$ for the census of Problem-solving Courts by State in the U.S.}
 \label{Fig:MsrsVrblty}
\end{figure}

Figure~\ref{Fig:MsrsVrblty} depicts the values of the four different measures of variability 
considered above for each territory in the dataset. 
The horizontal organization of the territories is presented in ascending order for the value 
obtained by $\sigma_\mathrm{L}$. 
An interesting fact can be addressed here: Although, globally, all variability 
measures present a growing tendency, locally, behaviour discrepancies are clearly visible. 
This can be seen comparing the smoothness of the purple line against the oscillation of the others. 
This implies that a direct comparison of the results among variability measures is not the best 
way to compare their performance.
This is another motivation to highlight that an analysis of robustness of variability measures under 
uncertainty is a better criterium to select the more accurate and meaningful measure of variability.

\begin{figure*}
\centering
 \begin{tabular}{ c c }
  \includegraphics[width=0.425\textwidth]{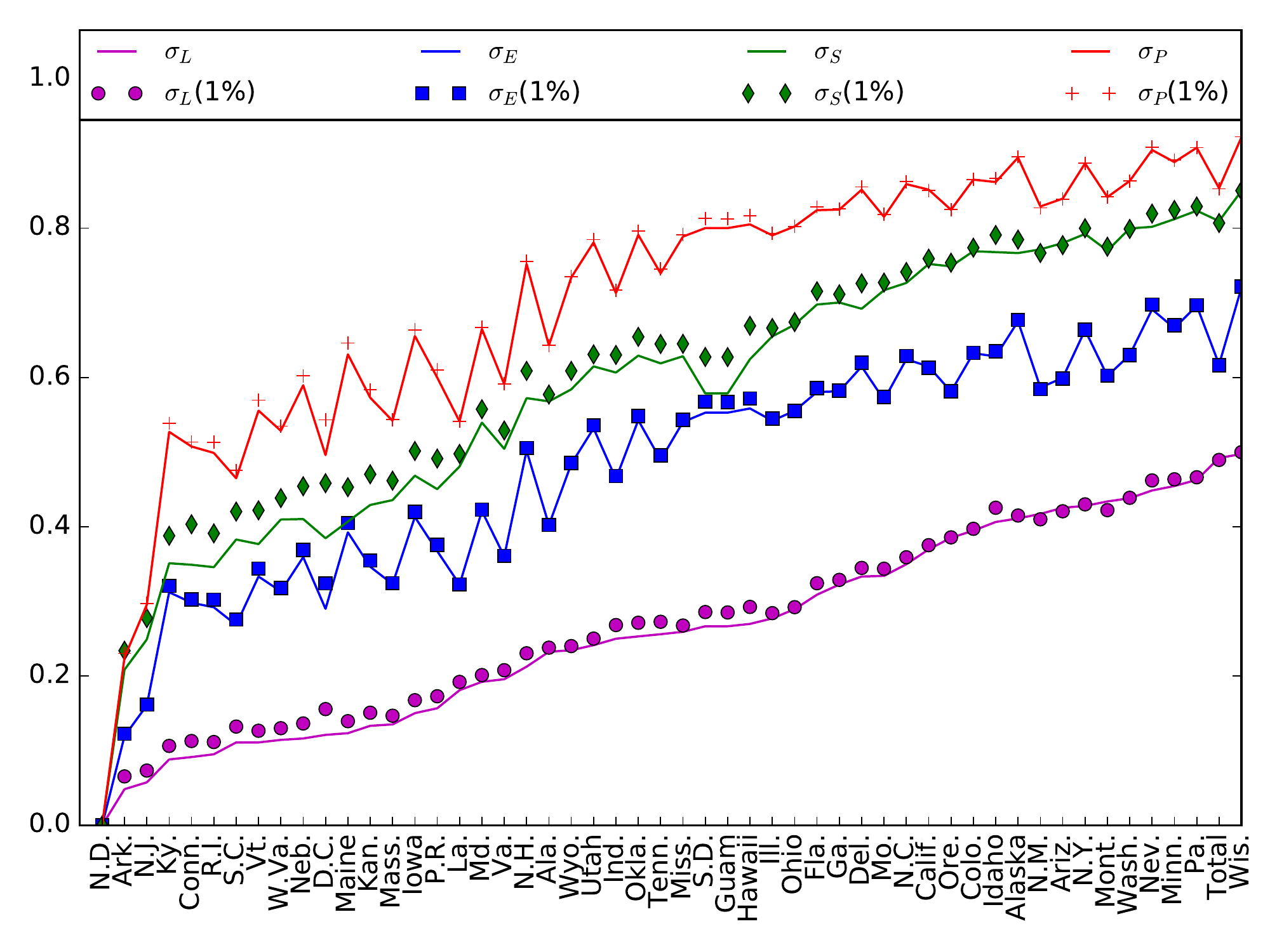} & \includegraphics[width=0.425\textwidth]{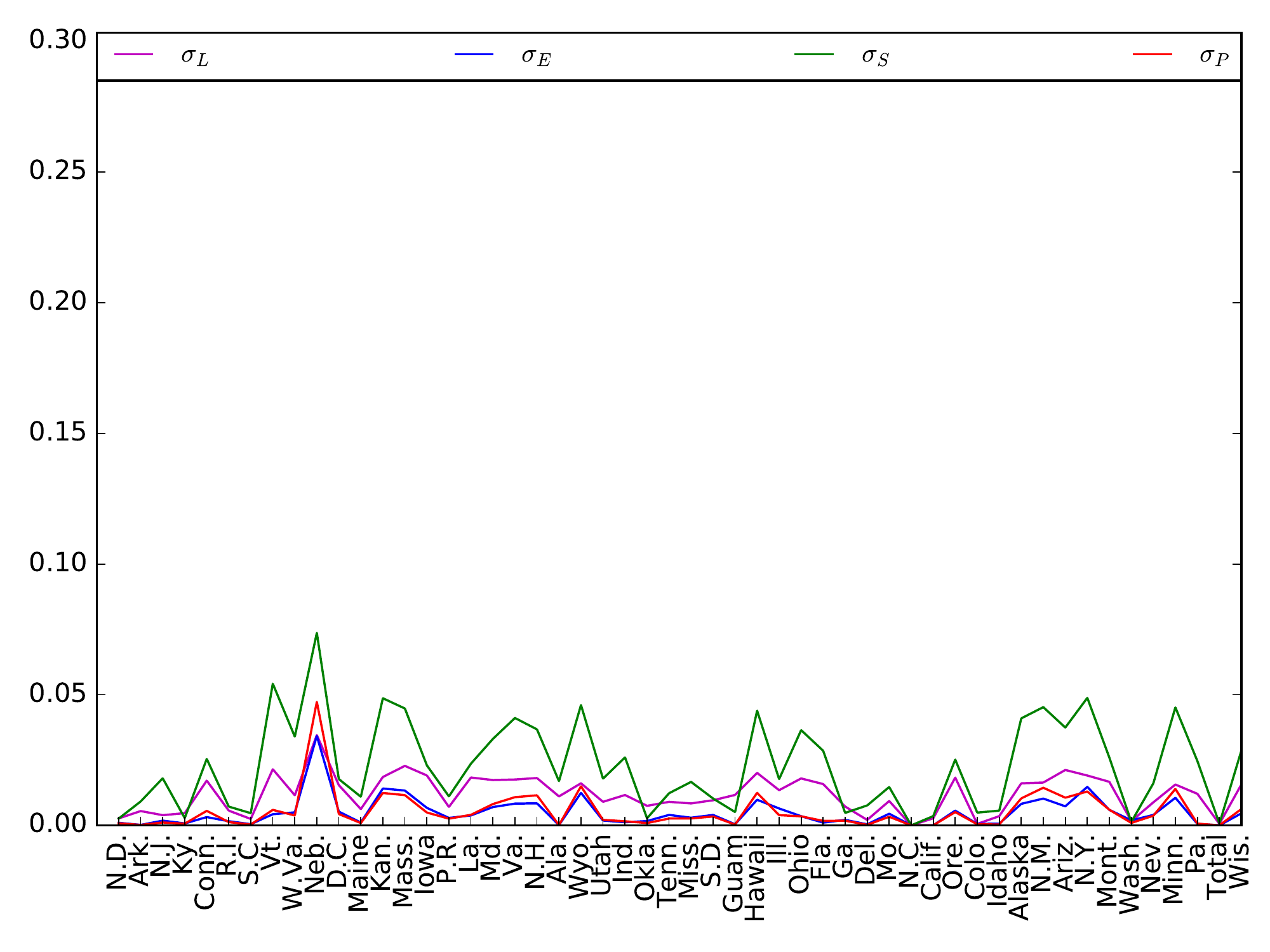} \\
  \includegraphics[width=0.425\textwidth]{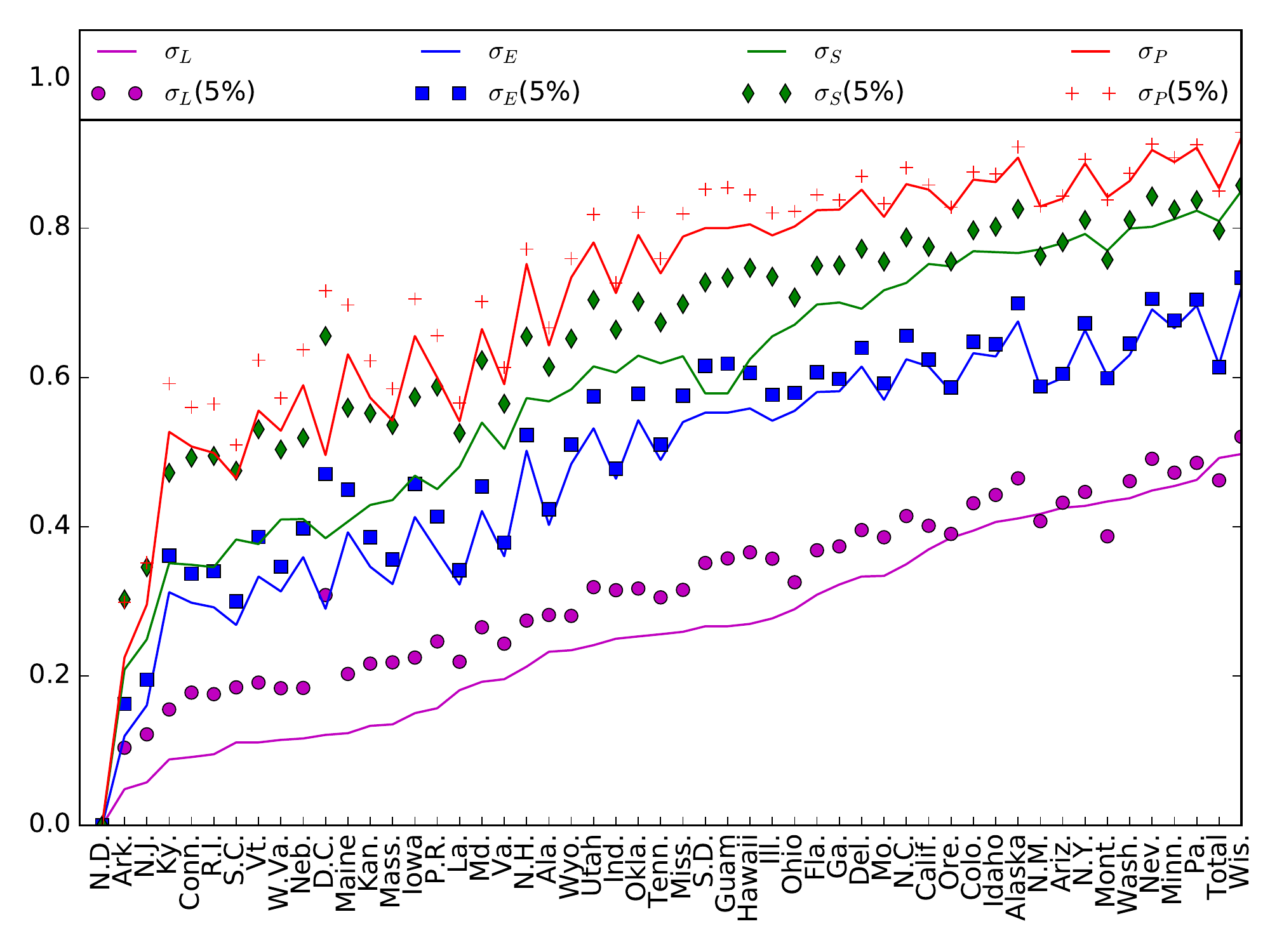} & \includegraphics[width=0.425\textwidth]{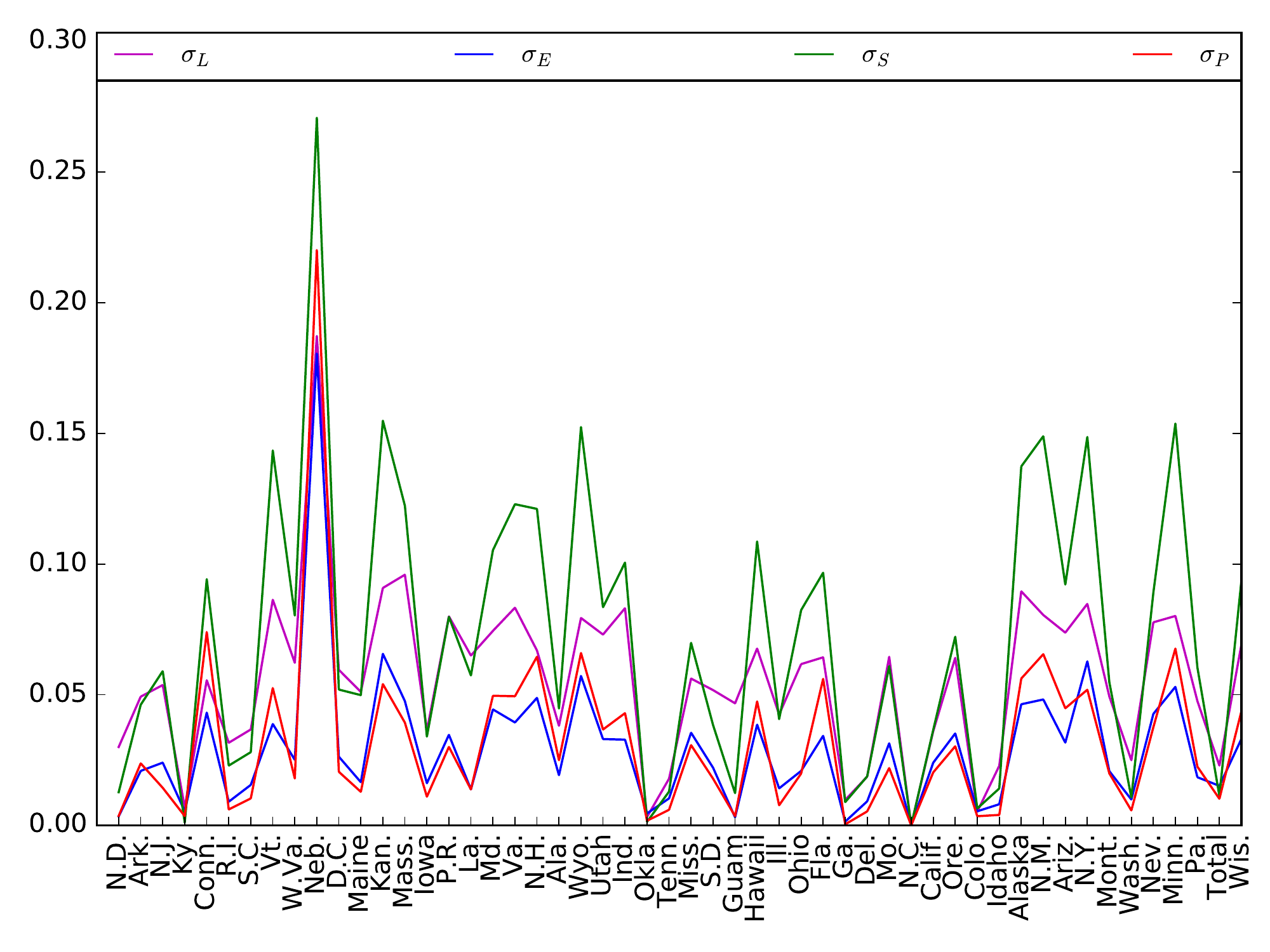} \\
\end{tabular}
\caption{Four different measures of variability ($\sigma_{\mathrm{L}}$, $\sigma_{\mathrm{E}}$, $\sigma_{\mathrm{S}}$ and 
$\sigma_{\mathrm{P}}$) of problem-solving courts by state in the U.S. 
For the upper panel the noise strength $1\%$ whereas for the lower panel it is $5\%$. 
{\bf Left Pannel} shows in a continuous line the measure of variability 
for the four cases and the result after introducing the noise. 
{\bf Right Pannel} shows the absolute value of the difference between the results 
obtained with noise and without introducing noise for each value of intensity. }
 \label{Fig:noise}
\end{figure*}
Figure~\ref{Fig:noise} depicts the results of the four different measures of variability considered 
in Fig.~\ref{Fig:MsrsVrblty} with the intensity of the noise $1\%$ (upper panel) and $5\%$ (lower panel) 
for each of the territories.
For the sake of comparison, the continuous curves, on the left-hand-side panel, depict 
the variability measure for U.S. states without noise whereas the results in the presence
of uncertainty are depicted by single markers in the figures (see caption of the figure for more 
details).
The right-hand panel shows the absolute value of the difference between the variability with and 
without noise for each value of noise intensity. 
It can be easily seen that the green line is higher than the other measures; therefore, 
$\sigma_{\mathrm{S}}$ can be interpreted as the most sensitive measure of noise, followed 
by $\sigma_ {\mathrm{L}}$ (purple line). 
From the picture is not easy to identify whether 
$\sigma_ {\mathrm{E}}$ (blue) or $\sigma_ {\mathrm{P}}$ (red) is more sensitive to the noise. 
Therefore, to quantitative measure the difference between the curves with and without 
noise, an measure equivalent to the standard deviation is considered for this case. 
Specifically, assume that the mean value will be given by the no-noise case whereas the 
noisy situations are to be understood as ``experimental'' data. 
The robustness against noise is then quantified by
\begin{equation}
 \label{eq:std_noise}
 \Phi_x(\theta)=\sqrt{\sum_{\alpha=1}^{\Lambda}[\sigma_x^{\alpha}(\theta)-\sigma_x^{\alpha}]^2},
\end{equation}
where $\theta$ represent the noise strength (e.g., $1\%$ or $5\%$), $\Lambda$ 
is the total number of territories and $x=\{\mathrm{L, E, S, P}\}$ labels one of the four possibilities 
of variability measure. 
The values of $ \Phi_x(\theta)$ can be found in Table~\ref{tab:std_noise}. 
\begin{table}
 \centering
 \begin{tabular}{| c | c | c | c | c |}\hline
      & $\Phi_\mathrm{L}$    & $\Phi_\mathrm{E}$    & $\Phi_\mathrm{S}$& $\Phi_\mathrm{P}$ \\ \hline
$1\%$ & $0.0138593$ & $0.0073292$ & $0.0283516$ & $0.0090894$ \\
$5\%$ & $0.0630688$ & $0.0382758$ & $0.0881197$ & $0.0438631$ \\ \hline
 \end{tabular}
 \caption{Results of the measurement of the dispersion of the values obtained for the different 
 territories without and with different values of noise, those results are obtained using the 
Eq.~(\ref{eq:std_noise}) }
 \label{tab:std_noise}
\end{table}
The values obtained for $\Phi_ {\mathrm{E}}$ are smaller than the others, so it can be concluded 
that $\sigma_ {\mathrm{E}}$ is the most robust measure of qualitative variation.  

\section{Variation of categories analysis}
\label{Sec:VrtnCtgrs}
Traditionally, the way to analyze how the reduction of categories affects the variability 
is by means of the Wilcox's proposal \cite{Wil73}, namely, by replacing by zero the 
proportion of the reduced category and then renormalize the proportion of the 
remaining categories. 
This proposal disregards the information of the reduced categories and assumes 
that the categories-to-be-reduced do not exist.
This formulation violates the basic postulates of information science as well as the second 
law of thermodynamics because entropy-like measures decrease when information decreases 
(randomness increases). 
Below, it is described how the Hilbert space formulation allows performing a reduction of 
categories while keeping the information provided by the reduced categories and consistent
with information science.
For simplicity, the reduction of categories analysis will be performed in the singly-excited 
manifold, so that the density operator reduces to
{\small
\begin{align}
\label{equ:dnstyoprtrSEM}
\rho=
 \bordermatrix{         & | 0 ... 0 ...0\rangle & | 1 0 ...0\rangle & | 0 ... 1_i ...0\rangle & | 0 ... 0 ...1_k\rangle \cr 
\langle 0 ... 0 ...0|   & 0                     & 0                 & 0                       & 0                                             \cr
\:\:\langle 1 0 ...0|   & 0                     & p_1               & 0                       & 0                                          \cr
\langle 0 ... 1_i ...0| & 0                     & 0                 & p_i                     & 0                                           \cr
\langle 0 ... 0 ...1_k| & 0                     & 0                 & 0                       & p_k                                     \cr
}.
\end{align}
}
For convenience, the case when none of the categories is chosen, element $| 0 ... 0 ...0\rangle$, 
was included in the definition in Eq.~(\ref{equ:dnstyoprtrSEM}) so that the density operator
has dimensions $(k+1)\times(k+1)$.
To perform the reduction of categories, the partial trace technique \cite{breuer2007theory} is applied 
over the $j^\mathrm{th}$-category to be reduced.
This procedure yields a new density operator, $\hat{\rho}_{\mathrm{r}_j}=\mathrm{tr}_j \hat{\rho}$, 
of dimension $k \times k$ that is to be replaced, e.g., in equations (\ref{eq:dispLop}-\ref{eq:dispPop})  to calculate the change in the variability. 
The normalization factor needs to be replaced appropriately, speficically, it is replaced in favor
of $\hat{\rho}_{\mathrm{r}_j,\mathrm{max}}=\mathrm{tr}_j \hat{\rho}_\mathrm{max}$.

Due to the intrinsic diagonal character of $\hat{\rho}$, the reduced density matrix $\hat{\rho}_{\mathrm{r}_j}$ 
will have all the information associated with the elements that it keeps and also information about 
the effects of the traced out elements (see, e.g., Sec.~2.2 in Ref.~\cite{breuer2007theory}). 
This characteristic is fundamental in applying the reduction of categories in a way consistent with 
information science \cite{breuer2007theory}.

The application of the partial trace for the present situation is straightforward.
Since $\rho$ is diagonal, performing the partial trace on the state $| 0 ... 1_j ...0\rangle$ yields 
a reduced density matrix of order $2^{k}$ with the same elements of the total density matrix $\rho$. 
The proportion $p_j$ associated with the $j^\mathrm{th}$-category switches to the none answer 
position, i.e., in the first input of the matrix. 
If the process is repeated over different states, the result is adding up the proportion associated 
with the traced element to the none answer position. 
Therefore, applying the partial trace over $n$ of the $k$ categories produces a reduced density 
operator $\hat{\rho}_{\mathrm{r}_{\{j,l,\ldots n\}}}$ of order $2^{k-n+1}$ given by
\begin{widetext}
\begin{align}
\label{eq:traced_matrix}
\rho_{\mathrm{r}_{\{j,l,\ldots n\}}}=  \bordermatrix{         & | 0 ... 0 ...0\rangle & | 1 0 ...0\rangle & | 0 ... 1_i ...0\rangle & | 0 ... 0 ...1_k\rangle \cr 
\langle 0 ... 0 ...0|   & p_j+p_l+\cdots+p_n    & 0                 & 0                       & 0                                              \cr
\:\:\langle 1 0 ...0|   & 0                     & p_1               & 0                       & 0                                              \cr
\langle 0 ... 1_i ...0| & 0                     & 0                 & p_i                     & 0                                              \cr
\langle 0 ... 0 ...1_k| & 0                     & 0                 & 0                       & p_k                                          \cr
}.
\end{align}
\end{widetext}
After applying the reduced density operator to the equations~(\ref{eq:dispLop}-\ref{eq:dispPop}), 
where $n$ category reduction is performed,
the following relatively simple expressions, in terms of the probabilities $\{p\}$, account for measures of 
variability 
\begin{widetext}
\begin{equation}
  \label{eq:dispLtraced}
 \sigma_\mathrm{L}=1-\frac{1}{2(k-n-1)}\sum_{\substack {b=1 \\ b\notin R}}^{k}\left(\left|\sum_{a\in R} p_a-p_b\right|+\sum_{\substack {b'=1 \\ b\notin R}}^{k}\left|p_b-p_{b'}\right|\right) 
\end{equation}

\begin{equation}
  \label{eq:dispStraced}
  \begin{split}
     \sigma_\mathrm{S}=1-\frac{1}{2(k-n-1)}\left(\left[\sum_{\substack {b=1 \\ b\notin R}}^{k}\left(\sum_{a\in R} p_a-p_b\right)^2\right]^{\frac{1}{2}} +
\left[\sum_{\substack {b,b'=1 \\ b,b'\notin R}}^{k}(p_b-p_{b'})^2\right]^{\frac{1}{2}}\right)
  \end{split}
\end{equation}

\begin{equation}
 \label{eq:dispSHAtraced}
 \sigma_\mathrm{S}=\frac{\left(\sum_{a\in R} p_a\right) \log_2 \left(\sum_{a\in R} p_a\right)+\sum_{\substack {b=1 \\ b\notin R}}^{k} p_b \log_2 p_b}{\log_2{(k-n)}} 
\end{equation}
\begin{equation}
 \label{eq:dispPURtraced}
 \sigma_\mathrm{P}=\frac{k-n}{k-n-1}\left[1-\left(\sum_{a\in R}p_a\right)^2-\sum_{\substack {b=1 \\ b\notin R}}^k p_b^2\right]
\end{equation}
\end{widetext}
where $R$ is the set of index associated with the reduced categories i.e. $R=\{j,l,...,n\}$. 
Not that that the number of categories have been reduced from $k$ to $k-n$. 
%
%

\begin{figure}
 \centering
 \begin{tabular}{c}
  \includegraphics[width=0.45\textwidth]{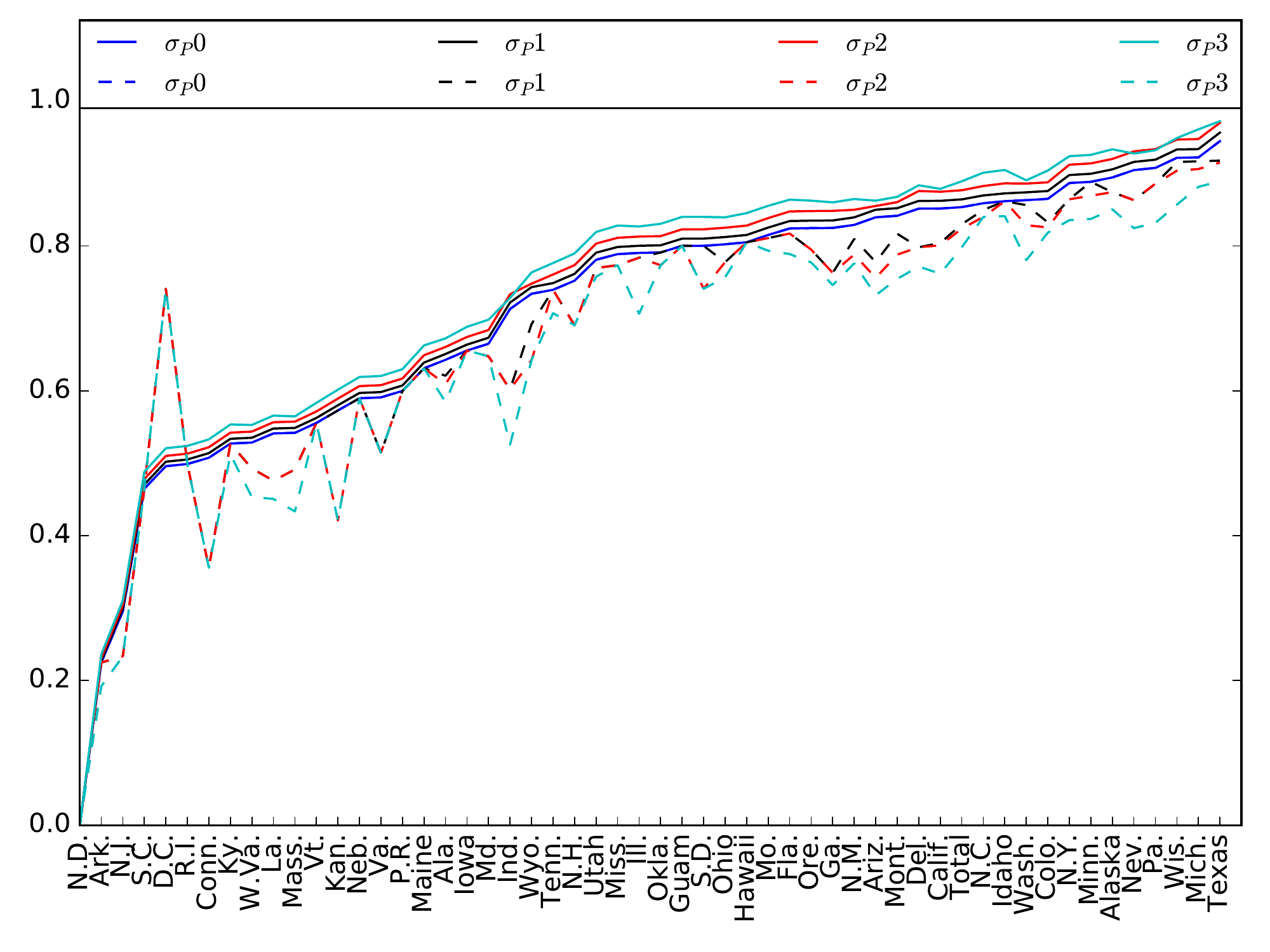} 
  \\
  \includegraphics[width=0.45\textwidth]{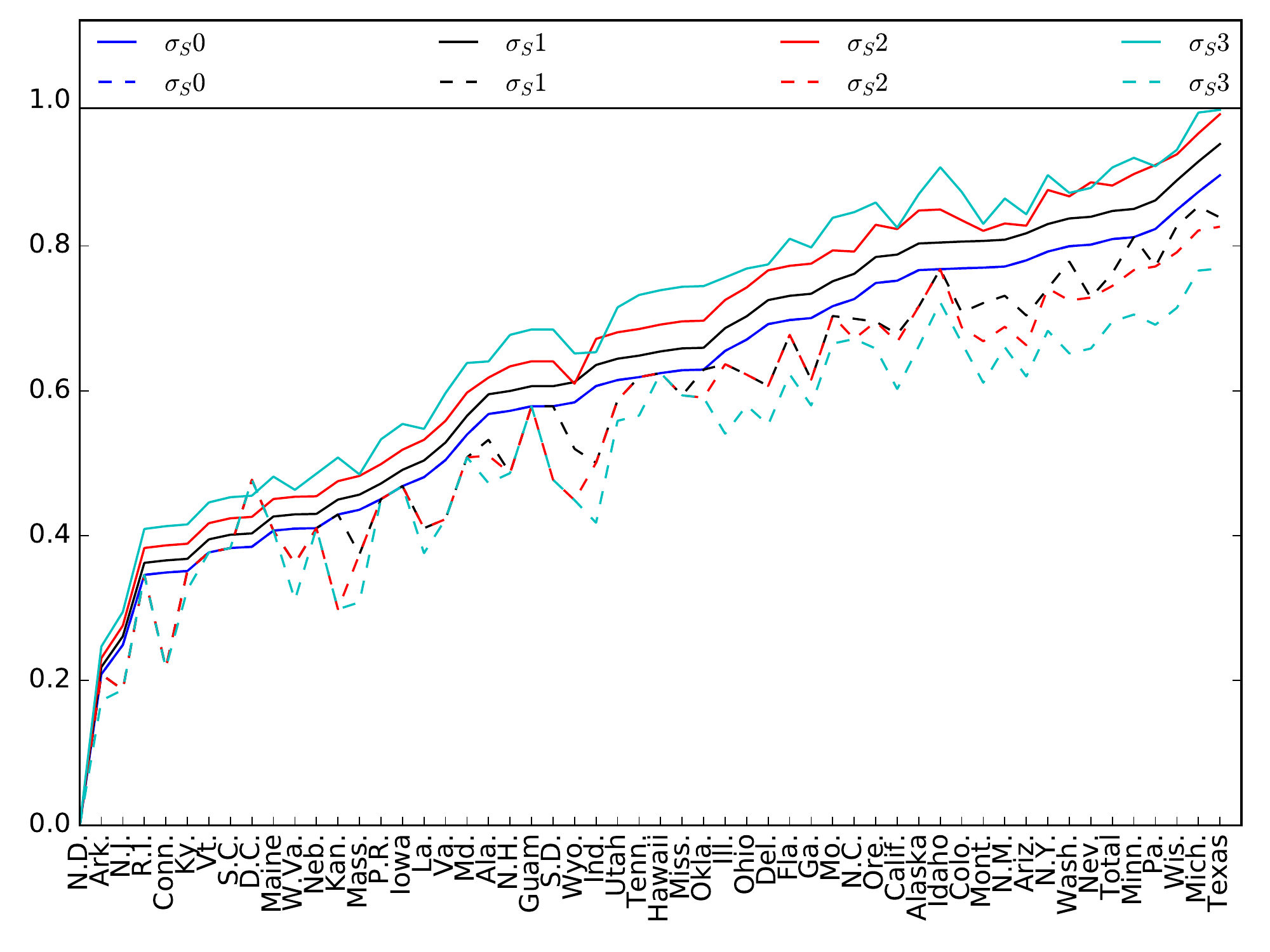}
 \end{tabular}
 \caption{Continuous lines depict the results for the measure of $\sigma_\mathrm{P}$ (top) 
 and $\sigma_\mathrm{S}$ (bottom) applying the reduction of categories using the partial trace 
 method, we also present the results in dashed line when the reduction is applied using the Wilcox 
 method. 
 The notation $\sigma$0,$\sigma$1, $\sigma$2, $\sigma$3 means that no category was reduced 
 and one, tow and three categories were reduced respectively.
}
  \label{Fig:RdctnCtgrs}
\end{figure}

Figure~\ref{Fig:RdctnCtgrs} depicts the results obtained for $\sigma_\mathrm{P}$ (top) and 
$\sigma_\mathrm{S}$ (bottom), applying the reduction of categories to the same dataset of the 
Problem-solving courts. 
From the initial ten categories scenario, the results of reduction of up to three categories are 
presented. 
Those results were obtained using the partial trace method (continuous line) and the Wilcox 
method (dashed line). 

The reduction of categories is applied from the right to the left in Table 1 of Ref.~\cite{SRK16}. 
As it can be seen from the figures, for all the territories in the U.S., the reduction of categories 
using the partial trace methods, implies an increase in the values of $\sigma_\mathrm{S}$ and 
$\sigma_\mathrm{P}$, conversely, a reduction of the values is obtained if the implemented 
method is the proposed by Wilcox. 
In terms of information, this is a very important issue, because the variability measures 
[Eqs.~(\ref{eq:dispSHAop}) and (\ref{eq:dispPop})] quantifies how the values between the 
categories are distributed, but in terms of information it can be understood as: if the value of variability 
approach to zero we have more information about organization between the categories and 
on the other hand, if the value approaches to 1 the lack of information information increases.
Hence, when the reduction of categories is applied, information is lost, then it is 
natural to expect that the value of the variability increases instead of decreasing.


No that the variability for  D.C. the increases when the categories are reduced using 
the Wilcox method; apparently, contradicting the conclusion above. 
This increase is due to the explicit value of the proportion of each reduced category. 
In this particular case, the first category reduced was \emph{other} and for D.C. has 
a very high proportion of $0.727$. 
Hence, ignoring that category, as in the Wilcox method, has a tremendous impact on the 
values of $\sigma_\mathrm{S}$ and $\sigma_\mathrm{P}$, otherwise, the method 
proposed by us keeps that information and thus do not produce a significant alteration 
in the curves. 

\section{Conclusions}
Two main contributions have been done with this research: (i) Introducing the Hilbert space 
formalism and its advantages to solve reduction of categories problem of  qualitative variables
and (ii) Proposing a robustness analysis as a methodology to choose the best qualitative 
variation for a dataset. In more detail:

(i) Using the Hilbert space formalism, the matter of reduction of categories, an issue unsolved 
since the late 70's is solved in a simple and very elegant way by performing the partial trace 
over the categories wanted to reduce. 
Importantly, this approach is consistent with information science.
This allows for the extension of analysis without the necessity of a constructing a new dataset 
which sometimes is one of the biggest problems faced by social scientists.
The formalism also allows the manipulation of datasets with simultaneous choice options.

(ii) A purely numerical methodology to choose the best variability measurement to implement 
in a dataset is new and provides a strong criteria if the dataset could have any bias or noise, 
like most of the real-datasets. 
The robustness against noise is a very important characteristic for indices, due to the fact that 
the intention of the index is to be useful and as general as possible.

The potencial of the present formalism goes beyond the application of physics concepts and 
tools to social sciences and reach the field of Artificial Intelligence by providing formal support
to the  One Hot Encoding Algorithm  \cite{MG16}.
%
\section{Acknowledgments}
Juan D. Botero was supported by Colciencias through a doctoral scholarship (Program
No. 727)

\bibliography{vqvPRX}


\end{document}